\documentclass{aa}
\usepackage{graphicx}

\newcommand{\Web}{{\tt Web}}
\newcommand{\GOF}{{\em GoF}}
\newcommand{\like}{{\cal L}}

\newcommand{\EfbP}{\hat{\delta T_{\rm fb}}}

\newcommand{\OmM}{\Omega_{\rm m}}
\newcommand{\Ho}{H_{\rm o}}
\newcommand{\Omb}{\Omega_{\rm b}}
\newcommand{\etab}{\eta_{10}}

\newcommand{\lamo}{\lambda_{\rm o}}
\newcommand{\Omk}{\Omega_\kappa}
\newcommand{\OmT}{\Omega_{\rm tot}}

\newcommand{\Da}{D_{\rm a}}
\newcommand{\da}{d_{\rm a}}

\begin{document}

%
   \title{Cosmological Constraints from the Cosmic Microwave Background}
%

        \titlerunning{Cosmological Constraints from the CMB ...} 

   \author{M.~Le~Dour$^1$, M.~Douspis$^1$, J.G.~Bartlett$^1$, 
                  A.~Blanchard$^{1,2}$} 

   \offprints{bartlett@ast.obs-mip.fr}

   \institute{$^1$ Observatoire Midi-Pyr\'en\'ees,
              14, ave. E. Belin,
              31400 Toulouse, FRANCE \\
              Unit\'e associ\'ee au CNRS, UMR 5572
             ({\tt http://www.omp.obs-mip.fr/omp})\\    
              $^2$ Universit\'e Louis Pasteur,
                   4, rue Blaise Pascal, 
                   67000 Strasbourg,
                  FRANCE\\
             }

   \date{April 2000}

   \maketitle

   \begin{abstract}
     Using an approximate likelihood method 
adapted to band--power estimates, we 
analyze the ensemble of first generation
cosmic microwave background anisotropy 
experiments to deduce constraints over
a six--dimensional parameter space describing
Inflation--generated adiabatic, scalar 
fluctuations.  The basic preferences
of simple Inflation scenarios are consistent
with the data set: flat geometries $(\OmT
\equiv 1-\Omk \sim 1)$
and a scale--invariant primeval spectrum
($n\sim 1$) are favored.  Models 
with significant negative curvature 
($\OmT < 0.7$) are eliminated, while 
constraints on postive curvature 
are less stringent.  Degeneracies
among the parameters prevent independent
determinations of the matter density
$\OmM$ and the cosmological constant
$\Lambda$, and the Hubble constant
$\Ho$ remains relatively unconstrained.
We also find that the height
of the first Doppler peak relative
to the amplitude suggested by data at
larger $l$ indicates 
a high baryon content ($\Omb h^2$), 
almost independently of 
the other parameters.  Besides the overall
qualitative advance expected of the next generation
experiments, their improved dipole calibrations
will be particularly useful for constraining the
peak height.
Our analysis includes a {\em Goodness--of--Fit}
statistic applicable to power estimates and
which indicates that the maximum
likelihood model provides an acceptable fit to 
the data set.

   \end{abstract}


\section{Introduction}

        The newest and perhaps most powerful tool in the 
cosmologist's toolbox are the temperature fluctuations
in the cosmic microwave background (CMB).  Their very
existence (e.g., Smoot et al. 1992) 
lends much credence to the general picture of
gravitational instability forming galaxies and the 
observed large--scale structure.  The last two decades
have witnessed the elaboration of this idea, with
numerical studies of gravitational growth permitting quantitative
comparison to actual survey data, and with the 
introduction of a physical mechanism -- namely,
Inflation -- for the creation of the initial perturbations.
Daring in scope, the resulting scenario would
encompass the evolution of the Universe from possibly
the Planck era to the present, explaining not only
the origin of the required density perturbations,
but also dispelling a handful of misgivings 
about the initial conditions of Big Bang model, such
as the impressive homogeneity on large scales
(Kolb \& Turner 1990; Peebles 1993; Peacock 1999)
Like any theory, it can never be proved; but it
may be tested.  And like any good theory,
it provides a physics, dependent on the validity
of the theory, specific enough to be used as a tool
to other ends:  it is now well appreciated that 
detailed study of the CMB fluctuations may
be applied to determine the fundamental 
parameters of the Big Bang model itself
(Bond et al. 1994; Knox 1995; Jungmann et al. 1996). 

        Our object with the present study is
to examine what may be learned from present
CMB data within the Inflationary context.
It should be noted at the outset that
there are other models contending to
explain the origin of density perturbations,
for example, defect models (Durrer 1999).  
Neither these nor any
other alternative shall be our concern in
the following, and it is important to 
emphasize that our results are therefore
limited to the Inflationary context.  
Very specifically, we shall be concerned only
with temperature fluctuations caused by
adiabatic, coherent and passively evolving
density (scalar) perturbations, dominated
by cold dark matter (CDM) (Bond \& Efstathiou 1984;
Vittorio \& Silk 1984).  Inflation
can generate isocurvature modes, but we 
ignore these in the following, along
with gravitational waves, the only non--scalar 
modes expected in Inflationary scenarios, 
and reionization.  All model predictions 
have been calculated using the CAMB Boltzmann 
code developed by Lewis, Challinor \& Lasenby (1999)
and built upon CMBFAST (Seljak \& Zaldarriaga 1996;
Zaldarriaga, Seljak \& Bertschinger 1998)
The reason for these restrictions
is of course that it is impossible to
explore the otherwise vast parameter space.

     Even in this seemingly restrained setting,
the physics of CMB temperature
anisotropies is quite rich, but it may 
nevertheless be boiled down to two regimes:
large scales where purely gravitational effects
operate [Sachs--Wolfe (SW) effect; Sachs \& Wolfe 1967],
and small scales, within the horizon at 
recombination where causal physics
plays its role.  In the latter regime, 
pressure of the coupled baryon--photon
fluid resists the inward pull of gravity
and therein establishes oscillating 
sound waves that will be observed in
the CMB fluctuation power spectrum
as a sequence of power peaks 
(so--called Doppler peaks; e.g., Hu \& Sugiyama 1996).  These 
are damped towards the smallest
scales by smoothing due to the
finite thickness of the last scattering
surface from which emanates the CMB.
The well defined aspect of these power
peaks owes to the coherent nature
and passive evolution of Inflation
generated perturbations; the loss
of both of these in defect models
has the effect of broadening or completely
smearing out the peaks (see, e.g,. Durrer 1999 
and references therein).
The very existence of such a series
of power peaks is thus a strong discriminator
between models.

     Besides the exact details of
the physics of Inflation,
the development of such perturbations 
depends, unsurprisingly, on the constituents of the 
primordial plasma in which they reside,
as well as on the metric background
in which they evolve.  This provides 
the link between the observable sky
temperature fluctuations and the
fundamental cosmological parameters.
Actual data from the first generation 
of CMB experiments\footnote{we use this
term to refer to those experiments, with the exception 
of COBE, which were not specifically designed for 
map making; our detailed list is given in Table 2 and 
discussed below.} already indicates the existence of the
first Doppler peak, a fact that leads to 
interesting and non--trivial conclusions, as noted
by many authors and as detailed below.

     In this {\em paper} we present our constraints
over a six--dimensional parameter space (see Table 1)
resulting from a large compilation of first
generation experiments (see Table 2).  Our
conclusions are based not on a traditional 
$\chi^2$ approach, but on methods developed
and presented elsewhere which attempt to
account for the non--Gaussian nature of 
power spectrum estimates as well as
information at times lost in simple
band--power estimates [Bartlett et al. 1999 (BDBL);
Douspis et al. 2000 (DBBL); Douspis, Bartlett \& 
Blanchard (DBB)]; we have tested
these methods against complete likelihood
analyses over subsets of the present
CMB data.  Our parameter space here 
is spanned by the quantities listed in Table 1, 
namely, the Hubble constant $\Ho$; the
total energy density $\OmT\equiv \OmM+\lamo$, where
$\OmM$ is the matter density and $\lamo\equiv \Lambda/3$;
the vacuum energy density parameter $\lamo$;
the baryon density, $\Omb$,  in terms of $\etab\equiv$
(baryon--to--photon number ratio)$\times 10^{10}$
($\Omb h^2 = 0.00366\etab$); and the primordial
scalar spectral parameters $n$ and $Q$, i.e.,
the slope and normalization.  Our data set is
listed in detail in Table 2. 
 
While some recent analyses have covered a larger parameter
space (e.g., Tegmark \& Zaldarriaga 2000), they are based 
on traditional $\chi$--squared methods. 
In a key work, Dodelson \& Knox (1999) applied approximate
likelihood methods similar to those employed here, although over
a more restricted range of parameters.  The present work 
thus extends the approximate likelihood approach to
a larger parameter space and includes an appropriate 
\GOF\ statistic.  On the other hand, we have not
considered the effects of calibration uncertainties;
the agreement between our results and those of 
previous authors supports the idea that these 
do not make a drastic difference to the final conclusions.

\begin{figure}\label{fig_powplot}
\begin{center}
\resizebox{\hsize}{!}{\includegraphics[angle=0,totalheight=8.4cm,
        width=8.cm]{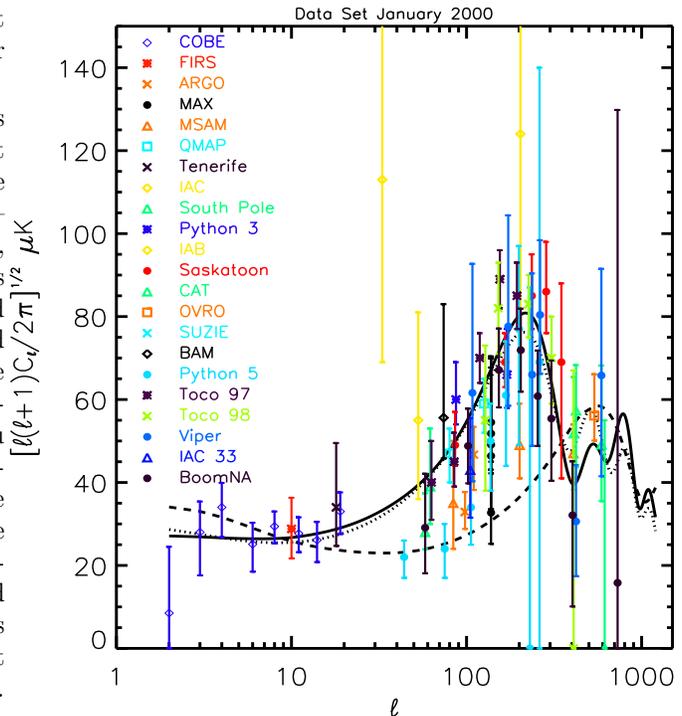}}
\end{center}
\caption{The power plane: measured flat--band power estimates and model 
spectra as a function of multipole.  The solid line shows
our best fit model (by approximate maximum likelihood) excluding
the Python V points (see text): 
$(\Ho,\OmT,\lamo,\Omb h^2,n,Q)=(60\; $km/s/Mpc,$
1.0,0.3,0.032,1.06,16.0\; \mu$K). 
Our \GOF\ statistic indicates that this is an acceptable
fit.  We also plot as the 
dotted line the best model with with a fixed $\Omb h^2=0.011$, 
falling $2\sigma$ below the prefered value: 
$(\Ho,\OmT,\lamo,\Omb h^2,n,Q)=(50\; $km/s/Mpc,$
1.1,0.8,0.011,1.00,19.0\; \mu$K).
The dashed line is an open model
with $\OmT=0.2$ and $\Ho=60$ km/s/Mpc, shown for illustration
and clearly excluded by the data.
}
\end{figure}    

     The {\em paper} develops as follows: we begin
in the next section with a presentation of 
our analysis methods; for the most part, this is a brief
review of work presented in BDBL, DBBL and DBB. 
Our parameter constraints are then given in Section
3, followed by our conclusions in Section 4.  The basic and 
robust result at this stage must be considered
as the conclusion that the spatial curvature is
close to zero and the primordial spectral
index close to its scale--invariant value of
unity.  The results of our adapted statistical analysis 
are thus in agreement with many 
others (Lineweaver et al. 1997; Bartlett et al.
1998ab; Bond \& Jaffe 1998; Efstathiou et al. 1998; 
Hancock et al. 1998; Lahav \& Bridle 1998; 
Lineweaver \& Barbosa 1998ab; Lineweaver 1998; Webster et al. 1998;
Lasenby et al. 1998; Dodelson \& Knox 1999; Melchiorri et al. 1999;
Tegmark \& Zaldarriaga 2000; Knox \& Page 2000) 
and confirm the two fundamental predictions
of the Inflationary model.

\section{Analysis Method}

     The experimental results of Table 2 are 
shown in Figure 1 as standard band--power 
estimates, together with several theoretical 
curves.  As for most current analyses, this will
be our starting point.  Simple as it may at first
appear, a correct statistical approach to these
data is in fact a non--trivial issue.  
The temptation is of course to apply a 
traditional $\chi^2$ minimization to the
ensemble of points and errors.  This, however,
is not strictly allowed, for these points,
being {\em power} estimates, are {\bf not} Gaussian
distributed.  In addition, the given errors are
usually computed from the the band--power
likelihood function and do {\bf not} therefore
necessarily represent the errors on the power estimate
(which a frequentist would argue must be found
by simulation).  In general,
the best statistical method to constrain parameters
with CMB data would be a likelihood analysis based on 
the original {\em pixel} values (or temperature
differences; in this paper, we use the term
`pixel' to refer to differences as well), because
this is guaranteed to use all relevant experimental
information.  Straightforward to construct
in the (present) context of Gaussian theories
as a multi--variant Gaussian in the
pixel values, whose covariance matrix depends
on the underlying model parameters (and noise characteristics),
the likelihood is in practice computationally
cumbersome due to the required matrix operations
(Bond, Jaffe \& Knox 1998; Borrill 1999ab; Kogut 1999).

\begin{table}[h]
\begin{center}
\begin{tabular}{|c|c|c|c|c|c|c|}
\hline
 & $\Ho$ (km/s/Mpc)&$\OmT$&$\lamo$&$\etab$&$n$&$Q\; \mu$K\\
\hline
\hline
Min. & 20  & 0.1   & 0.0   & 1.11  & 0.70  & 10.0 \\
\hline
Max. & 100 & 2.0   & 1.0   & 10.66 & 1.42  & 25.0 \\
\hline
step & 10  & 0.1   & 0.1   & 1.91  & 0.06  & 1.5  \\
\hline
\hline
\end{tabular}
\end{center}
\caption{Parameter space explored:} 
\vspace{-0.3cm}
$\OmT \equiv 1-\Omk$, where $\Omk$ is the curvature parameter \\
$\lamo \equiv \Lambda/3$ \\
$\etab \equiv$ (baryon number density)/(photon number density)\\
\hspace*{0.9cm} $\times 10^{10}$ \hspace*{0.2cm} (Note: $\Omb h^2 
= 0.00366\etab$)\\
$n\equiv$ primeval spectral index\\
$Q\equiv \sqrt{(5/4\pi) C_2}$ \\
\end{table}

     This difficulty has motivated us (BDBL; DBBL; DBB) 
and others (Bond, Jaffe \& Knox 1998; Wandelt, Hivon \& G\'orski 1998) 
to propose methods based on power estimates and aimed
at reproducing as far as possible a complete
likelihood analysis.  Computation time
is greatly reduced by working in the power plane
(Bond, Jaffe \& Knox 1998; Tegmark 1997),
due to the much smaller number of data elements, 
but all proposed methods must be {\em benched} against
the ultimate goal of recovering the likelihood
results. In DBBL we evaluated at length the viability of 
several approximate methods by comparison to a complete
likelihood analysis.  We found that the traditional 
$\chi^2$ method over power estimates (such as shown
in Figure 1) is subject to bias.  Other approaches
based on approximate band--power likelihood functions
fare better, but still do not always fully reproduce
the complete likelihood results.  As discussed in 
DBBL, this may be traced to  
relevant experimental information lost by the simple
band--power representation of an experiment;
we demonstrated this explicitly by showing
that some MAX and Saskatoon data are sensitive
to the local slope of the power spectrum.  
In other words, the observations constrain
not only an in--band power, but also a local
effective spectral slope.  This information is
correctly incorporated by a complete likelihood
analysis, but obviously missed by any method
based solely on the band--power estimates of 
Figure 1.  

\begin{figure}\label{fig_gof}
\begin{center}
\resizebox{\hsize}{!}{\includegraphics[angle=0,totalheight=6.cm,
        width=6.cm]{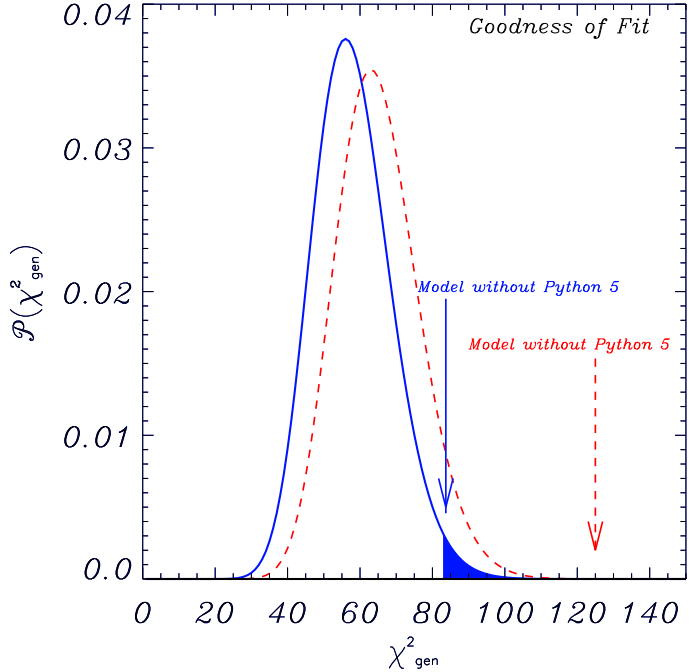}}
\end{center}
\caption{\GOF\ statistic: shown are the distribution and
observed value of our generalized $\chi^2$ for the fit
without Python V (blue, solid distribution and arrow),
and for the fit including the Python V points (dashed,
red distribution and arrow).  The fit is reasonable in
the former case, but unacceptable in the latter (see text).}
\end{figure}

     These comments motivate an approach
in which all relevant experimental information
is first identified by a set of parameters
that are constrained by the data.  A 
general likelihood function may then be
constructed over this parameter space 
using the original pixel set; this need be
done only {\em once}.  Expressing a general
power spectrum by these same parameters
then permits one to assign a likelihood
value to any model by extrapolation of
the pre--calculated likelihood function.
This value incorporates all
pertinent information and should be the
best possible approximation to the exact
likelihood.  The gain is that one 
manipulates the likelihood function with
the full pixel set only once, and then 
simply interpolates over a much reduced set
of parameters.  Once the likelihood function
has been calculated, the technique is hardly
more complicated than the traditional $\chi^2$
employed for its facility.
Unfortunately, the information needed 
for a general application of this technique to all 
of the current experimental results is not readily available.
Because the number of points we are able to treat 
in this fashion is thus small, we elect to 
analyze the entire data set with the band--power
approximation developed in BDBL.

        We examine a set of Inflationary models over
the parameter space spanned by $(\Ho,\OmT,\lamo,\etab,
n,Q)$, with respective ranges and step sizes
given in Table 1.  The likelihood of each model
is calculated as just described, and the
best model is found by maximizing the
likelihood function over the explored space.
We present our results as a series of 
two--dimensional contour plots of the likelihood
projected onto various parameter planes.  
The contours are defined in the full six--dimensional
space with values 
of $\Delta\log(\like)=1, 4$ (dashed, in green),
motivated as the 1 and 2 $\sigma$ contours 
of a Gaussian distribution 
when projected onto one of the axes, and of
$\Delta\log(\like)=2.3, 6.17$ (solid, in red), motivated by a 
Gaussian distribution with two degrees of
freedom.  Since the likelihood is not actually Gaussian, 
the confidence percentages associated with our 
contours are not exactly 1 and 2 ``$\sigma$''; the technique is
however standard practice.  
Our final results shown in the following figures
have been obtained by {\em excluding} the
Python V points.  As discussed below, 
this is because the inclusion of this data set
leads to a poor \GOF for the entire class of models
considered.  On formal grounds we would then be lead to 
reject the models, while on a technical note we worry that 
likelihood contours for a poor best--fit may give 
misleading constraints.  Perhaps somewhat arbitrarly,
we thus proceed in our analysis without these data.
This difficulty would in part be alleviated by
a complete treatment of calibration errors.

\begin{figure}\label{fig_LO}
\begin{center}
\resizebox{\hsize}{!}{\includegraphics[angle=0,totalheight=9cm,
        width=10cm]{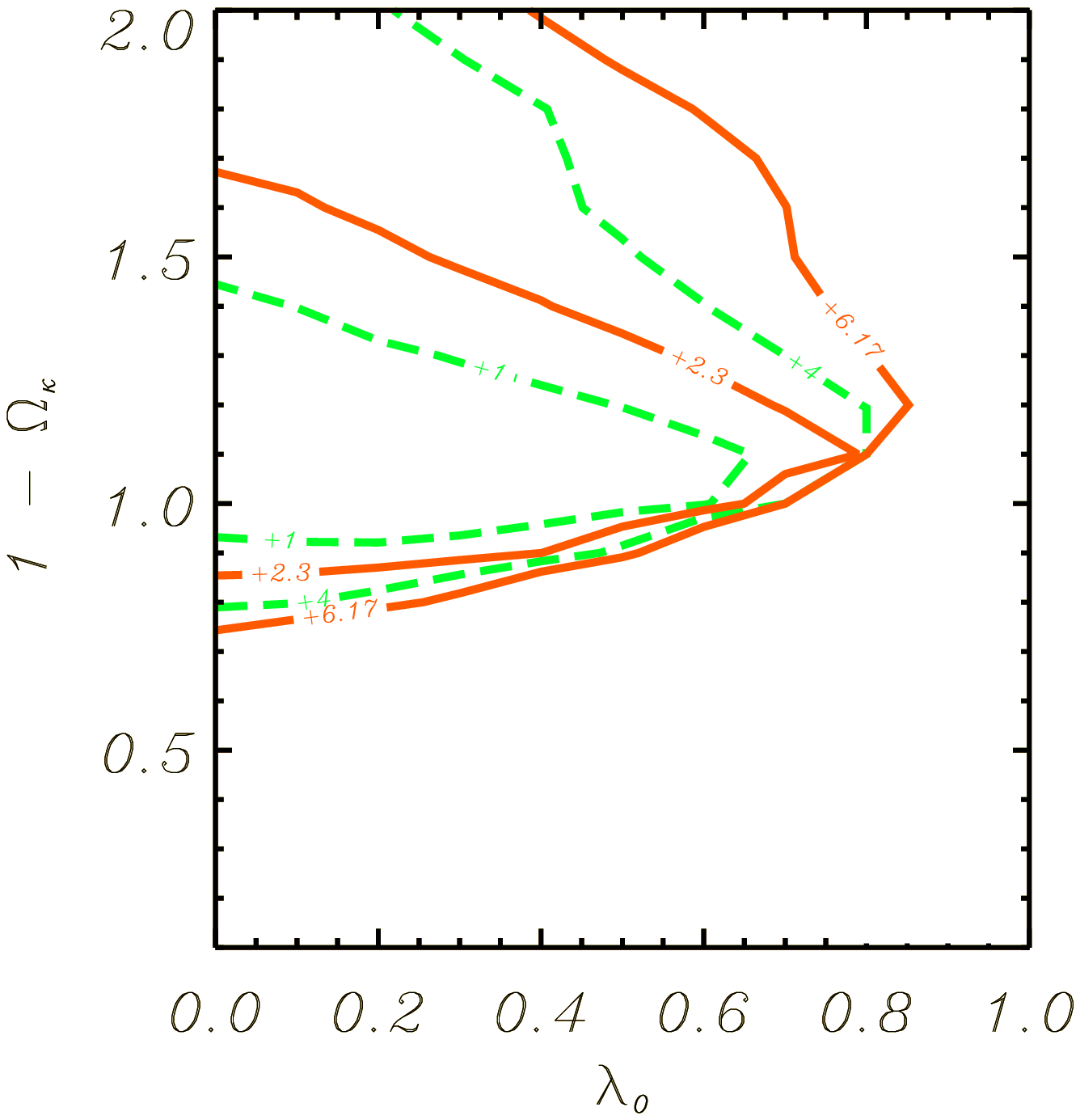}}
\end{center}
\caption{Constraints in the $(\OmT,\lamo)$--plane, where 
the other parameters have been projected out; 
$1=\OmT+\Omk=\OmM+\lamo+\Omk$, and 
$\lamo\equiv \Lambda/3$.  Hyperbolic models are strongly
ruled out, but the constraints are less stringent on 
spherical geometries.  The degeneracy between $\OmM$
and $\lamo$ preventing independent determinations of the matter
and vacuum contributions is manifest as the horizontal orientation
of the contours.}
\end{figure}

    Another aspect of our analysis
not incorporated, as far as we are aware,
in previous work is the application of an adequate 
{\em goodness--of--fit} statistic (\GOF).  Once the 
best model, i.e., the most likely model, has been found, 
one is obliged to evaluate the quality of its description
of the data.  As for the likelihood function
itself, our situation is complicated by the
fact that the power estimates shown in Figure 1
are not Gaussian distributed variables; 
in particular, a traditional $\chi^2$ \GOF\ 
statistic is inadequate for the task.  In 
DBB we proposed a \GOF\ statistic
readily applicable, if necessarily approximate,
to band--power estimates. One requires
{\em distribution} of these power estimates,
$\EfbP$, for a given, underlying model;
this distribution is not the same as the 
band--power likelihood function (frequentist
point--of--view).  Remarkably, we found 
in DBB that
the same parameters introduced in BDBL 
to approximate the band--power
likelihood function could be re--employed
in a slightly different fashion to yield
the distribution of the power estimator.
The technique was tested with Monte Carlo simulations
of experiments for which we performed 
complete likelihood analyses; details may
be found in DBB.  The important point
is that with just the best power estimate
and a confidence interval, we may construct
an approximation to the complete
distribution of the power estimate
from an experiment and, hence, a 
\GOF\ statistic for the data ensemble
shown in Figure 1.

\section{Results}

     Although perhaps at first glance the 
observational situation shown in Figure 1
seems confused, in fact the first Doppler peak
would appear clearly detected.  Several different
experiments viewing different regions of sky
on this scale, such as BOOMERanG, Python V, Saskatoon,
Toco and QMAP, all indicate the presence of 
a rise in power and, in some cases, a hint
of the subsequent fall--off, which is also
supported by other experiments at higher
$l$.  This is not to say that all the data
follow exactly the same party line, one example
being the rather low MSAM points around the peak,
but one could argue that the general trend favors
the presence of a rise in power over the scales
expected for first Doppler peak of Inflationary 
scenarios.  

     The way to quantify these statements
is by fitting a model to the data and examining
its \GOF\ statistic.  Over the parameter space
explored and excluding the Python V points for the moment,
the data identify the model with 
$(\Ho,\OmT,\lamo,\Omb h^2,n,Q) = (60\; {\rm km/s/Mpc},1.0,0.3,0.032,1.06,16.0\; \mu{\rm K})$
as the best fit; the corresponding model spectrum
is shown in Figure 1 as 
the solid curve.  
The quality of the fit may be judged from
the distribution of our \GOF\ statistic
(referred to as a generalized $\chi^2_{\rm gen}$)
as shown in Figure 2.  Assuming the data
come from the adopted model, a value of 
$\chi^2_{\rm gen}$ as big or larger than 
the observed value (indicated by the heavy 
arrow) occurs with a probability
of 0.014 (i.e., 1.4\% of the time; indicated by the
shaded area under the curve).  
Admittedly, this is a little
low, but it does give a rather satisfying 
numerical representation of the impression given by 
the data (it seems {\em reasonable}).  
Although perhaps the fit is marginal, we certainly 
do not find the result sufficiently conclusive
to eliminate the entire class of models from consideration.
We are further comforted in this direction
by the fact that the \GOF\ is dominated by
only a few outliers (see DBB).  We thus do
not hesitate to accept the model and
move on to see what constraints the data provide
over the parameter space considered. 
We remark in passing that a (incorrect) standard $\chi^2$ 
statistic is even less kind to the model with a value
only 0.0016 (i.e., 0.16\%) probable.  Keeping 
the Python V points in the analysis yields
a very poor best--fit model, quantified by a value of  
$\chi^2_{\rm gen}$ only $2\times 10^{-5}$ probable (marked on 
the figure by the dashed (green) curve and
arrow).  This is the reason 
for which we choose to exclude Python V 
in the following; thus, all our final results quoted 
hereafter and shown in the figures exclude this data set.
Fully aware that such a procedure 
should always be undertaken only with
caution (perhaps the fluctuations are not
Gaussian, for example), we nevertheless feel
that this is the most constructive approach 
at present.  In some sense this is 
a purely formal argument, but we do worry about 
the interpretation of the contours 
from a likelihood with a poor \GOF.
In any case, the best--fit model
is only slightly changed by inclusion of 
Python V: $(\Ho,\OmT,\lamo,\Omb h^2,n,Q) = 
(60\; {\rm km/s/Mpc},1.0,0.5,0.032,1.12,13.0\; \mu{\rm K})$. 
For comparison, we also calculated the \GOF\ for 
the so--called ``concordance model'' 
$(\Ho,\OmT,\lamo,\Omb h^2,n,Q) = 
(65\; {\rm km/s/Mpc},1.0,0.3,0.018,1,20.0\; \mu{\rm K})$:
we find a \GOF\ 0.1\% probable.

\begin{figure}\label{fig_nO}
\begin{center}
\resizebox{\hsize}{!}{\includegraphics[angle=0,totalheight=9cm,
        width=10cm]{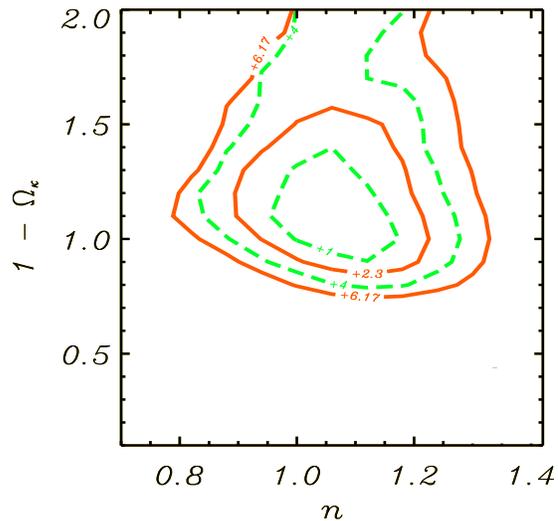}}
\end{center}
\caption{Constraints in the $(\OmT,n)$--plane.  We see
that the fundamental expectations of simple Inflation models
remain consistent with the data, namely, a flat 
geometry ($\Omk\sim 0$) and a scale--invariant, primeval 
density perturbation power spectrum ($n\sim 1$).
}
\end{figure}

     The presence of the first Doppler peak
in the data permits the elaboration of
non--trivial constraints over our parameter
space.  Within our present context
of adiabatic Inflation--generated perturbations,
this peak appears on the physical scale of 
the horizon at the moment of recombination,
$\sim \Ho^{-1}\sqrt{\OmM}$.  As the
distance to the last scattering surface 
is also proportional to $\Ho^{-1}$ 
($\Da = \Ho^{-1} \da(\OmM,\lamo)$) the
Hubble constant has relatively
little influence on the projected 
angular scale of the peak; rather,
$\OmM$ and, most notably, the curvature
of space (light ray focusing) 
control this observable scale (Blanchard 1984). 
For this reason, one should expect
that the most robust result coming
out of the present data set would be 
constraints in the $(\OmT,\lamo)$--plane,
as shown in Figure 3.  The remarkable 
conclusion is that a flat universe is 
prefered and that, in particular, models
with negative curvature (low $\OmT$) are 
eliminated.
On the other hand, the data place
only weak constraints on the value
of the cosmological constant.  This
degeneracy between $\OmM$ and $\lamo$ is
consistent with the expectation that
one constrains instead their combination 
defining the quantity $\sqrt{\OmM}/\da(\OmM,\lamo)$.

     If spatial flatness may be considered
as one of the motivating principals and a key
``prediction'' of the overall Inflation paradigm,
then another is certainly the form of the
primordial density perturbation spectrum, $n$.
Figure 4 shows our constraints in the 
$(\OmT,n)$--plane.  The two most simple
``predictions'' of Inflation are rapidly
evaluated on this diagram, and we see
that the model fares quite well:  zero
spatial curvature ($\OmT=1-\Omk=1$) 
and the spectral index of a scale--invariant 
spectrum ($n=1$) both fall within the inner contour.

    As already mentioned, the best--fit model
appears acceptable (marginally) according to our 
\GOF\ statistic.  Since our statistic is based
on the assumption of Gaussianity, this in particular
implies that the data are consistent with 
{\em Gaussian anisotropies}, as also expected in
the simplest Inflationary scenarios.
The general, overall conclusion from the first
generation CMB anisotropy experiments must then
be their coherence with the rudimentary concepts
of Inflation.

\begin{figure}\label{fig_HE}
\begin{center}
\resizebox{\hsize}{!}{\includegraphics[angle=0,totalheight=9cm,
        width=10.cm]{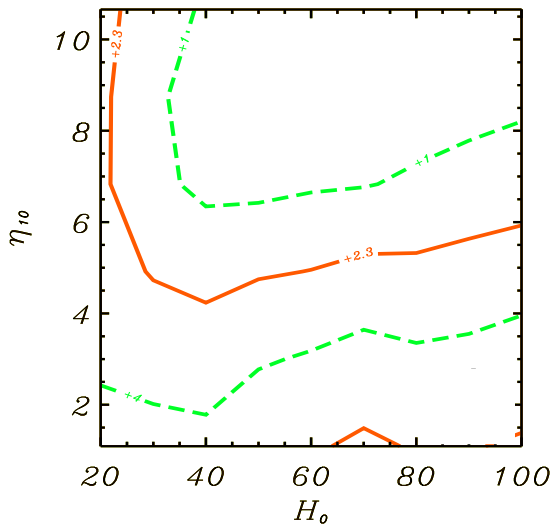}}
\end{center}
\caption{Constraints in the $(\etab,\Ho)$--plane.
We find that high values of $\etab$ are favored by
the present data set (see text).  This appears to
be primarily due to the height of the first Doppler
peak and to the fact that some data 
(e.g., BOOMERanG N.A. and Viper) indicate a 
rapid fall--off before the second Doppler peak.
For illustration, we also
plot in Figure 1 a model with $\etab= 3.0$  
(other parameters chosen to maximize
the likelihood for this value of $\etab$).
Note that the low D/H abundances in some 
QSO systems would correspond to 
$\etab=5$ (Tytler et al. 2000).
}
\end{figure}

     One surprise of our analysis concerns 
the baryon density, $\Omb$.  As seen in Figure 5,
the data indicate extremely high values of $\Omb h^2$, 
and this almost independently of the value of the
other parameters.  So--called low D/H values observed 
in some QSO absorption systems yield  $\etab\sim 5$ (Tytler et
al. 2000), while here the CMB data prefer
even higher baryon densities; although within 
``$2\sigma$'' the two remain consistent.
We believe the origin of this result to be 
related to the relative height of the first 
Doppler peak.  Some data, such as Toco 97 \& 98 and
Saskatoon, suggest a high peak followed by 
a deep trough; low power at higher $l$ is also 
supported by other experiments, like Viper.  For
illustration, we show as the dotted line in
Figure 1 the best--fit model with a fixed $\Omb h^2=0.011$.
The essential differences between this model and the overall
best--fit model are (in the fomer case) a lower first peak 
and the absence of 
a deep trough before the appearance of the second peak.
One important aspect (besides the anticipated large improvement
in overall data quality) of the next generation 
mapping experiments (e.g., 
Archeops\footnote{{\tt http://www-crtbt.polycnrs-gre.fr/archeops/\\general.html}},
BOOMERanG\footnote{{\tt http://astro.caltech.edu/\~lgg/boom/boom.html}},
and 
MAXIMA\footnote{{\tt http://cfpa.berkeley.edu/group/cmb/gen.html}})
is their ability to calibrate on the CMB dipole.  
This new calibration method, plus the fact that
the entire first peak will be covered by a single
instrument, should help to reduce 
the uncertainty surrounding the peak heights.
As we have just argued, this is particularly
important for determination of such quantities
as the baryon density, and it will be an
important test of the present result
favoring a high baryon density.

\section{Conclusions}

     Our purpose in this {\em paper} has been to
analyze the ensemble of first generation CMB
anisotropy experiments to see what conclusions may
be drawn concerning certain fundamental cosmological 
parameters from the CMB data {\em alone}.  
Our approach employs approximate
likelihood methods that are adapted to power estimates,
and which have been detailed elsewhere (BDBL, 
DBBL and DBB).  Our primary conclusions are
that a flat geometry ($\Omk\sim 0$ or $\OmT\sim 1$) and
a scale--invariant primeval spectrum ($n\sim1$) are favored,
while strongly hyperbolic models are 
ruled out with high significance -- in short, 
Inflation remains a good theory.  
Specifically, the best--fit model parameters are
$(\Ho,\OmT,\lamo,\Omb h^2,n,Q)=(60\; {\rm km/s/Mpc},1.0,0.3,0.032,1.06,16.0\; \mu{\rm K})$.  Our
analysis includes a \GOF\ statistic that
indicates that this model, and therefore the entire class (Inflation with 
adiabatic, scalar perturbations and without re--ionization) 
provides an acceptable description of the data.
Many authors have recently
explored these issues (Lineweaver et al. 1997; Bartlett et al.
1998ab; Bond \& Jaffe 1998; Efstathiou et al. 1998; 
Hancock et al. 1998; Lahav \& Bridle 1998; 
Lineweaver \& Barbosa 1998ab; Lineweaver 1998; Webster et al. 1998;
Lasenby et al. 1998; Dodelson \& Knox 1999; Melchiorri et al. 1999;
Tegmark \& Zaldarriaga 2000; Knox \& Page 2000), with various
combinations of the present data set, and most
would agree with these conclusions. 
The extensive spectral coverage 
(in multipole order $l$) over the first, and perhaps
second Doppler peaks, and the low noise expected of the 
next generation instruments should qualitatively change 
the confidence in and precision of this kind 
of study.

     Very little may be said about $\Ho$ or about
the relative contributions of matter, $\OmM$, and
vacuum, $\lamo$, to the total energy density,
the latter due to a well--known degeneracy 
when considering CMB data alone.  One must 
turn to other observational constraints, coming from, 
for example, cluster evolution (Oukbir \& Blanchard 1992), 
cluster baryon fractions (White et al. 1993), 
SNIa Hubble diagrams (Reiss et al. 1998; Perlmutter et al. 1999),
weak cosmic shear (Blandford et al. 1991;
Mellier 1999), etc..., 
to eliminate such degeneracies (so--called ``cosmic
complementarity'', Eisenstein, Hu \& Tegmark 1999).  
Many of the authors listed above
have included such constraints in their analysis; our
own work along these lines is left to a forthcoming paper.  

     A surprising note is the very high baryon densities 
that we have found are prefered by the data set:
$\etab\sim 8.9$ or $\Omb h^2\sim 0.032$.  This is
even higher than the values indicated by the
``low'' D/H values found in several QSO absorption
systems (Tytler et al. 2000), although within ``$2\sigma$'' (a little
more for high values if $\Ho$; see Figure 5) the 
results are consistent.  It remains to be seen
if this intriguing result bears the scrutiny 
of the next generation experiments.   
In particular, their ability to significantly reduce 
overall calibration uncertainty by using the CMB 
dipole will be crucial to such issues.  

\begin{table*}[h]
\begin{center}
\begin{tabular}{|c|c|c|c|}
\hline
Experiment&$\delta T \pm \sigma (\mu K)$&$\ell_{eff}$&reference\\
\hline
\hline
ARGO 1 &$32.9^{+4.8}_{-4.1}$&98&Ratra et al. 1999\\
\hline
ARGO 2 &$46.7^{+9.5}_{-12.1}$&98&Masi et al. 1996\\
\hline
BAM&$55.6^{+27.4}_{-9.8}$&74&Tucker et al. 1997\\
\hline
BOOM NA&$29.1^{+13.0}_{-11.0}$&58&Mauskopf et al. 1999\\
\hline
BOOM NA&$48.8^{+9.0}_{-9.0}$&102&Mauskopf et al. 1999\\
\hline
BOOM NA&$67.1^{+10.0}_{-9.0}$&153&Mauskopf et al. 1999\\
\hline
BOOM NA&$71.9^{+10.0}_{-10.0}$&204&Mauskopf et al. 1999\\
\hline
BOOM NA&$60.8^{+11.0}_{-12.0}$&255&Mauskopf et al. 1999\\
\hline
BOOM NA&$55.4^{+14.0}_{-15.0}$&305&Mauskopf et al. 1999\\
\hline
BOOM NA&$32.1^{+13.0}_{-22.0}$&403&Mauskopf et al. 1999\\
\hline
BOOM NA&$15.8^{+114.}_{-15.8}$&729&Mauskopf et al. 1999\\
\hline
CAT 1a&$51.8^{+13.6}_{-13.6}$&410&Scott et al. 1996\\
\hline
CAT 1b&$49.1^{+19.1}_{-13.6}$&590&Scott et al. 1996\\
\hline
CAT 2a&$57.3^{+11.}_{-13.6}$&422&Baker et al. 1997\\
\hline
CAT 2b&$0.0^{+55.0}_{-0.0}$&615&Baker et al. 1997\\
\hline
COBE 1&$8.5^{+16.}_{-8.5}$&2&Tegmark \& Hamilton 1997\\
\hline
COBE 2&$28.0^{+7.4}_{-10.4}$&3&Tegmark \& Hamilton 1997\\
\hline
COBE 3&$34.0^{+5.9}_{-7.2}$&4&Tegmark \& Hamilton 1997\\
\hline
COBE 4&$25.1^{+5.2}_{-6.6}$&6&Tegmark \& Hamilton 1997\\
\hline
COBE 5 &$29.4^{+3.6}_{-4.1}$&8&Tegmark \& Hamilton 1997\\
\hline
COBE 6 &$27.7^{+3.9 }_{-4.5}$&11&Tegmark \& Hamilton 1997\\
\hline
COBE 7 &$26.1^{+4.4}_{-5.3}$&14&Tegmark \& Hamilton 1997\\
\hline
COBE 8 &$33.0^{+4.6}_{-5.4}$&19&Tegmark \& Hamilton 1997\\
\hline
FIRS &$29.4 ^{+7.8}_{-7.7}$&10&Ganga et al. 1994\\
\hline
IAB   &$124.0^{+60.}_{-60.}$&203&Piccirillo et al. 1993\\
\hline
IAC 1 &$113.0^{+49.}_{-44.}$&33&Femenia 1998\\
\hline
IAC 2 &$55.0^{+27.}_{-28.}$&53&Femenia 1998\\
\hline
IAC 33&$43.0^{+12.5}_{-11.5}$&105 &Dicker et al. 1999\\
\hline
MAX GUM&$54.5^{+16.4}_{-10.9}$&138&Tanaka et al. 1996\\
\hline
MAX HR&$27.9^{+11.5}_{-4.7}$&130& Our analysis \\
\hline
MAX ID&$52.7^{+32.1}_{-10.4}$&120& Our analysis \\
\hline
MAX PH&$72.9^{+30.6}_{-10.7}$&131& Our analysis \\
\hline
MAX SH&$82.0^{+53.0}_{-15.0}$&121& Our analysis \\
\hline
MSAM 1&$35.0^{+15.0}_{-11.0}$&84&Wilson et al. 2000\\
\hline
MSAM 1&$49.0^{+10.0}_{-8.0}$&201&Wison et al. 2000\\
\hline
MSAM 1&$47.0^{+7.0}_{-6.0}$&407&Wilson et al. 2000\\
\hline
OVRO&$56.1^{+8.5}_{-6.6}$&537&Leitch et al. 2000\\
\hline
Pyth A&$60.0^{+9.0}_{-5.0}$&87&Platt et al. 1997\\
\hline
Pyth B&$66.0^{+11.}_{-9.0}$&170&Platt et al. 1997\\
\hline
Python V&$22.0^{+4.0}_{-5.0}$&44&Coble 1999 thesis\\
\hline
Python V&$24.0^{+6.0}_{-7.0}$&75&Coble 1999 thesis\\
\hline
Python V&$34.0^{+7.0}_{-9.0}$&106&Coble 1999 thesis\\
\hline
Python V&$50.0^{+9.0}_{-12.0}$&137&Coble 1999 thesis\\
\hline
Python V&$61.0^{+13.0}_{-17.0}$&168&Coble 1999 thesis\\
\hline
Python V&$77.0^{+20.0}_{-28.0}$&199&Coble 1999 thesis\\
\hline
Python V&$69.0^{+71.0}_{-69.0}$&261&Coble 1999 thesis\\
\hline
Python V&$0.0^{+87.0}_{-0.0}$&230&Coble 1999 thesis\\
\hline
QMap,K1&$47.0^{+6.0}_{-7.0}$&80&de Oliveira-Costa 1998\\
\hline
QMap,K2&$59.0^{+6.0}_{-7.0}$&126&de Oliveira-Costa 1998\\
\hline
Sask 1&$51.5^{+8.4}_{-5.3}$&86&Netterfield et al. 1997\\
\hline
Sask 2&$72.5^{+7.4}_{-6.3}$&166&Netterfield et al. 1997\\
\hline
Sask 3&$89.3^{+10.5}_{-8.4}$&236&Netterfield et al. 1997\\
\hline
Sask 4&$90.3^{+12.6}_{-10.5}$&285&Netterfield et al. 1997\\
\hline
Sask 5&$72.5^{+20.}_{-29.4}$&348&Netterfield et al. 1997\\
\hline
SP91   &$28.0^{+9.5}_{-6.7}$&58&Gundersen et al. 1995\\
\hline
SP94   &$36.3^{+13.6}_{-6.1}$&62&Gundersen et al. 1995\\
\hline
SuZIE&$0.0^{+44.0}_{-0.0}$&2340&Ganga et al. 1997\\
\hline
Tenerife &$34.0^{+15.5}_{-9.3}$&18&Guti\'errez et al. 2000\\
\hline
\hline
\end{tabular}
\end{center}
\end{table*}

\begin{table*}[h]
\begin{center}
\begin{tabular}{|c|c|c|c|}
\hline
Experiment&$\delta T \pm \sigma (\mu K)$&$\ell_{eff}$&reference\\
\hline
\hline
\hline
Toco 97&$40.0^{+10.0}_{-9.0}$&63 &Torbet et al. 1999\\
\hline
Toco 97&$45.0^{+7.0}_{-6.0}$&85 &Torbet et al. 1999\\
\hline
Toco 97&$70.0^{+6.0}_{-6.0}$&119 &Torbet et al. 1999\\
\hline
Toco 97&$89.0^{+7.0}_{-7.0}$&155 &Torbet et al. 1999\\
\hline
Toco 97&$85.0^{+8.0}_{-8.0}$&194&Torbet et al. 1999\\
\hline
Toco 98&$55.0^{+18.0}_{-17.0}$&128 &Miller et al. 1999\\
\hline
Toco 98&$82.0^{+11.0}_{-11.0}$&152 &Miller et al. 1999\\
\hline
Toco 98&$83.0^{+7.0}_{-8.0}$&226 &Miller et al. 1999\\
\hline
Toco 98&$70.0^{+10.0}_{-11.0}$&306 &Miller et al. 1999\\
\hline
Toco 98&$0.0^{+67.0}_{-0.0}$&409&Miller et al. 1999\\
\hline
Viper&$61.6^{+31.1}_{-21.3}$&108 &Peterson et al. 2000\\
\hline
Viper&$77.6^{+26.8}_{-19.1}$&173 &Peterson et al. 2000\\
\hline
Viper&$66.0^{+24.4}_{-17.2}$&237 &Peterson et al. 2000\\
\hline
Viper&$80.4^{+18.0}_{-14.2}$&263 &Peterson et al. 2000\\
\hline
Viper&$30.6^{+13.6}_{-13.2}$&422&Peterson et al. 2000\\
\hline
Viper&$65.8^{+25.7}_{-24.3}$&589&Peterson et al. 2000\\
\hline
\hline
\end{tabular}
\end{center}
\caption{First generation CMB data set listed in alphabetical order.
This information may also be found on our \Web\ site
({\tt http://webast.ast.obs-mip.fr/cosmo/CMB/}).  
Our final analysis
excludes the Python V results (see text).}
\end{table*}



\end{document}